\documentclass[fleqn]{article}

\usepackage{amsmath,amssymb}
\usepackage{graphicx}
\usepackage[noblocks]{authblk}
\usepackage[top=0.75in, bottom=0.75in, left=0.75in, right=0.75in, dvips]{geometry}
\usepackage{hyperref}
\usepackage[small]{caption}

\begin{document}

\title{{\sf On approximating two distributions from a single complex-valued function}}

\author[1]{W. D.~Flanders} 
\author[2]{G.~Japaridze}
\affil[1]{ 
Emory University, Atlanta GA, 30327, USA}
\affil[2]{Clark Atlanta University, Atlanta GA, 30314, USA}

\maketitle

\begin{abstract}
We consider the problem of approximating two, possibly unrelated probability distributions from a single complex-valued function $\psi$ and its Fourier transform. We show that this problem always has a solution within a specified degree of accuracy, provided the distributions satisfy the necessary regularity conditions. We describe the algorithm and construction of $\psi$ and provide examples of approximating several pairs of distributions using the algorithm. 
\end{abstract}

\section{Introduction}
In this paper we consider the problem of reconstructing two different probability distributions from a single complex-valued function with a specified degree of accuracy. This problem was suggested by the interpretation of the wave function $\Psi$ which is a solution of the Schr\"{o}dinger equation. As is well known $\Psi^{\star}(x)\Psi(x)$ is the probability density in a position space and $\widehat{\Psi}^{\star}(p)\widehat{\Psi}(p)$ is the corresponding probability density in a momentum space, where $\Psi^{\star}(x)$ is the complex conjugate of $\Psi(x)$ and $\widehat{\Psi}(p)$ is the Fourier transform of $\Psi(x)$ \cite{quant}.
We raise the related question as to whether {\it any} two distributions can be approximated within a specified error from a single function by calculating its modulus or that of its Fourier transform,  calculations like those used in Quantum Mechanics.

Our answer to this question is affirmative.

We prove this assertion by constructing a complex-valued function $\psi$ which approximates two given distributions within a given error, provided the two distributions satisfy some regularity requirements specified below. In particular, we show that the cumulative distributions can be approximated pointwise using the indefinite integral of the modulus of $\psi$ or of $\widehat{\psi}$.

The paper is organized as follows. In the next section we list assumptions, conventions and notation. In section 3 we first introduce an expression for evaluating the error in the approximation of the two given distributions from the modulus of $\psi$ or its Fourier transform. We then show how to construct this function. In the next section we list 4 pairs of distributions approximated using construction described in section 3.
In section 5 we discuss the method, limitations and possible generalizations. In the appendix we sketch a proof that the function $\psi$ approximates any two given distributions on subintervals as claimed and a proof that the cumulative distributions can be approximated pointwise (uniformly on closed intervals) using the indefinite integral of the modulus of $\psi$ or of $\widehat{\psi}$.

\section{Definitions, conventions and assumptions}
\subsection{Assumptions}
\noindent \begin{enumerate}
 \item $X$ and $P$ are any two variables; $f_X(x)$ is the probability density function for $X$ with corresponding cumulative distribution $F_X(x)$; $ f_P(p)$ is the probability density function for $P$ with corresponding cumulative distribution $F_P(p)$
  \item $ f_X(x)$ and $f_P(p)$
 have finite variances and they (and their square roots) are continuously differentiable with Fourier transforms in $L^2$ and in $L^1$ (for definitions of $L^{1}$ and $L^{2}$ see \cite{rudin}; throughout, we denote the norm in $L^{2}$ as $\| \|_{2}$)
 \item for every $\varepsilon >0$, there exist $X_b$ and $P_b$ such that: i) for all $x \in  [-X_b, X_b] \ $ and for all $p \in  [-P_b, P_b] \ $, $f_X(x)> $ 0 and $f_P(p)>$ 0; and, ii) $F_X(-X_b)< \varepsilon/16$, $F_P(-P_b)< \varepsilon /16$, $F_X(X_b)> 1- \varepsilon /16$, and $F_P(P_b)> 1 - \varepsilon /16$.
\end{enumerate}

\subsection{Definitions and Notations}
\begin{enumerate}
 \item Let $n$ be any integer, set $J= 2^n$, $\triangle x = X_b/J$, and $\triangle p=P_b/J$, where $X_b$ and $P_b$ are given in assumption 3; if necessary, increase $n$ (and/or $P_b$) so that $F_P(P_b-\triangle p/2 )> 1 - \varepsilon /16$
 ; $2J$ is the number of intervals of width $\triangle x$ and $\triangle p$ in [$-X_b, \, X_b$] and 
 [$-P_b-\triangle p/2$, $P_b-\triangle p/2$]
 \item $s$ is a number defined in the Construction below; it indicates the number of subintervals into which each interval $\triangle x$ is divided. It is used to control the size of the error in the approximations
 \item $t=s$/($\triangle x$$\triangle p$)
 \item $j$ and $k$ are integers, between $-sJ$, and $sJ-1$; $l$, $m$ and $m'$  are integers between $-J$, and $J-1$ or $J$, unless noted otherwise
 \item $\phi_s(x)$ is a function of the form used in the "sinc approximation"\cite{sinc}, \cite{sinc2}:
\[\phi_s(x)= \sum^{sJ-1}_{j=-sJ} \sqrt {f_X(j \triangle x /s)}\, \textrm{sinc}\biggl(  \frac{x-j \triangle x /s}{\triangle x /s}  \biggr), \]
where
\[
\textrm{sinc}(x)\,\equiv\,{\sin(\pi x)\over \pi x}
\]
 \item $Q_x$  =  max $f_X(x)$, for $x \in [-X_b,\, X_b]$; $P_x$ = min  $f_X(x)$ $>$ 0, for $x \in [-X_b,\, X_b]$; $P_p$ = min  $f_P(p) >$0, for $p \in [-P_b-\triangle p/2 ,\, P_b]$; $Q'_x$ = max  $df_X(x)/dx$, for $x \in [-X_b,\, X_b]$
 \item $a_j =  [ \frac{\triangle x}{s} f_X(j\triangle x/s)]^{1/2} $,  for $j = -sJ, \,-sJ+1, \cdots, sJ-1$
 \item $pr_{x;j,z}$ = $ \sum^{j+z}_{m=j} a_m^2$, for $0 \leq z \leq s\cdot J-j $
  \item $pr_{p;l}$ = $\int^{(l+1)\triangle p}_{l\triangle p}\,f_P(z)dz $ for $-J \leq l \leq J-1$
   \item $ \hat{f}(.)$ denotes the Fourier transform of a function $f(.)$, $f^*(.)$ denotes the complex conjugate of a function $f(.)$
 \item $M_j$ is a function, defined recursively in the Construction, mapping the integers $j\in $ \{ $-sJ, - s J+1,..., s J-1$ \} to the numbers \{$- sJ /\triangle x, -s(J-1)/ \triangle x,..., s(J-1) /\triangle x,\, sJ/ \triangle x$\}
 \item $m_j=M_j \cdot \triangle x/s$; $m_j$ is an integer
 \item $S_l \equiv $ \{ $j: j \in \{-sJ, -sJ+1,...,sJ-1 $\} , and $M_j = l \cdot s/ \triangle x $
\} for $l=-J, -J+1,..., J$. Equivalently, $S_l$= $f_M^{-1}(l\cdot s/\triangle x)$, where $f_M^{-1}( .)$ is the inverse mapping
\end{enumerate}
\section{Construction}
\noindent Given $f_X(x), \, f_P(p)$, $\varepsilon $ and $n$, and $X_b$ and $P_b$ with the properties described in Assumptions we construct a complex-valued function $\psi$ which satisfies the following two inequalities
\begin{equation}
\label{claim1}
       \sum^{J-1}_{l=-J}\left | {\int^{^{(l+{1 \over{2}})\triangle p}}_{_{(l-{1 \over{2}})\triangle p}}
       \biggl(t \cdot \widehat{\psi}(qt)   \cdot \widehat{\psi}^*(qt) \,-\,f_{P}(q)\biggr)dq} \right | < \varepsilon
\end{equation}
and
\begin{equation}
\label{claim2}
       \sum^{J-1}_{l=-J} \left| {\int^{(l+1)\triangle x}_{l\triangle x}\, \biggl(\psi(x)   \cdot
       \psi^*(x)\,-\,f_{X}(x)\biggr)dx}\right | <\varepsilon
\end{equation}
where $t$ is a scale parameter defined as
\begin{equation}
\label{scale}
t\,=\,{s\over \triangle x \triangle p}
\end{equation}
and $s$  is defined to be the smallest integer so that inequalities (\ref{c1})-(\ref{c4}) are satisfied:

 \begin{equation}
 \label{c1}
 \frac{\triangle x}{s} Q_x < {\varepsilon \over 4J}
 \end{equation}
 \begin{equation}
 \label{c2}
 4\biggl[\frac{\triangle x}{s} Q_x \biggr]^{1/2} \biggl(3+2\cdot log(2sJ) \biggr)^3 < {\varepsilon\over 4(2J+1)^4}
 \end{equation}
 \begin{equation}
 \label{c3}
    \| \phi_s(x)-\sqrt {f_X(x)} \|_2 < \varepsilon /8(2J+1)^4
    \end{equation}
     \begin{equation}
 \label{c6}
     \frac{(\triangle x)^2}{s} J|Q'_x|\sqrt{\frac{Q_x}{P_x}} \biggl(2+2\cdot log(2sJ)\biggr)^2  \leq {\varepsilon\over 4(2J+1)^4}
     \end{equation}
     \begin{eqnarray}
   \label{c4}
  && \int^{X_b}_{-X_b}\,f_X(x)dx - \sum^{sJ-1}_{j=-sJ} f_X(j\triangle x/s)\frac{\triangle x}{s}\cr
&&=\sum^{sJ-1}_{j=-sJ} \int^{(j+1)\triangle x/s}_{j\triangle x/s} f_X(x)dx
    - \sum^{sJ-1}_{j=-sJ} f_X(j\triangle x/s)\frac{\triangle x}{s}\cr &&\leq {\varepsilon \over 8}
    \end{eqnarray}
Many of the functions already defined, such as $\phi _s(x)$, or to be defined, such as $\psi (x)$ and $M_j$, depend on $s$ and $n$. For brevity we suppress explicit notation of that dependency.

The next step in the construction of $\psi$ is to define subintervals. First we split the interval  $[-X_b,\, X_b]$ into $2J$ intervals of length $\triangle x$: $[l\triangle x, \,(l+1)\triangle x]$, $l=-J,\,-J+1,\,\cdots, J-1$. Next we further divide each interval  $[l\triangle x, \,(l+1)\triangle x]$ into $s$ subintervals and number them from $j =  -s J, - s J
+1,...,$ to $sJ-1$. Define ($j = -s J,\, - s J +1,\cdots s J-1$)
\begin{equation}
\label{aa}
a_j = \biggl [ \frac{\triangle x}{s} f_X\biggl({j\triangle x\over s}\biggr)\biggr]^{1/2}
 \end{equation}
We continue the construction by introducing the function $M_j$ which maps the integers $j\in $ \{ $-sJ, - s J+1,..., s J-1$ \} to the numbers \{$- sJ /\triangle x, -s(J-1)/ \triangle x,..., s(J-1) /\triangle x,\, sJ/ \triangle x$\}, using the following recursive algorithm: initialize $j$ to $-sJ$ and $l$ to $-J$. Then proceed with the following steps:
\begin{description}
\item[{\bf I}] Set $pr_{x;j,z} = 0$ for $z\,=\,-\,1$ and  set $pr_{x;j,z} = \sum^{j+z}_{m=j} a_m^2$ for each integer\newline $z\,\in\, $ \{ $0,\, 1\cdots sJ-j$ \}, and set\\
          $pr_{p;l}$ = $\int^{(l+{1 \over{2}})\triangle p}_{(l-{1 \over{2}})\triangle p}\,f_P(z)dz $ for $-J \leq l \leq J-1 $
\item[{\bf II}] If an integer $z$ exists between $-1$ and $sJ-j$ such that the inequalities $0\leq pr_{p;l}  - pr_{x;j,z} \leq \varepsilon /(2(2J+1))$,  and  $pr_{x;j,z+1} - pr_{p;l}  > 0$ hold,
then set $k = z$; otherwise, set $k = sJ - j$
\item[{\bf III}] If $k \geq 0$ then set all $M_j,..., M_{j+k}$ equal to $ls/\triangle x$
\item[{\bf IV}] Set $j=j+k+1$ and $l=l+1$; if $j = sJ$, stop, otherwise go to step {\bf V}
\item[{\bf V}]If $l +1 = J$, set $M_j, M_j+1,..., M_{sJ-1}$ all to $sJ/\triangle x$ and stop; otherwise return to step {\bf I}.
\end{description}
Step {\bf II} always returns a value of $k$. The iterative algorithm continues until stopping, either at step {\bf IV} or {\bf V}.

Properties of $M_j$ that will be used subsequently include: for each  $j \in {-sJ,..., sJ-1}$, the mapping $M_j$ is a non-decreasing function of $j$; each $M_j\cdot s/\triangle x$ is an integer (i.e. $M_j\cdot \triangle x/s = m_j$ for some integer $m_j \in {-J,..., J}$ ).

Using this definition of $M_j$, we can now define $\psi (x)$ appearing in equations (\ref{claim1})-(\ref{claim2}) as:
\begin{eqnarray}
\label{phidef}
 & & \psi (x)=\sqrt {\frac{s}{\triangle x}} \sum^{sJ-1}_{j=-sJ} a_j \,\textrm{sinc}\biggl(  \frac{x-j \triangle x /s}{\triangle x /s}  \biggr)\,e^{2i\pi M_{j}x}=\cr\cr
         & &\sum^{sJ-1}_{j=-sJ} \sqrt {f_X\biggl({j\triangle x\over s}\biggr)}\, \textrm{sinc}\biggl(  \frac{x-j \triangle x /s}{\triangle x /s}  \biggr)\,e^{2i\pi M_{j}x}
\end{eqnarray}
Apart from $\exp(2i\pi M_{j}x)$ and the $sJ$th term in the series (\ref{phidef}), the function $\psi (x)$ represents
the Whittaker-Shannon interpolation formula  \cite{whit}-\cite{shanon} for $\sqrt {f_X(x)}$.

In addition to approximations given in (\ref{claim1})-(\ref{claim2}) we also prove uniform approximations to the cumulative distributions: let us define
\begin{eqnarray}
\label{claimm2}
F_P(p) &=&  \int^{p}_{-P_{b}-\triangle p/2} f_P(q)dq \cr
       \widetilde{F}_P(p) &=& \int^{p}_{-P_{b}-\triangle p/2} t \cdot \widehat{\psi}(qt)   \cdot \widehat{\psi}^*(qt)dq
\end{eqnarray}
and
\begin{eqnarray}
\label{claimm1}
F_X(x) &=&  \int^{x}_{-X_{b}} f_X(z)dz \cr
       \widetilde{F}_X(x) &=& \int^{x}_{-X_{b}}   \psi(z)   \cdot \psi^*(z)dz
\end{eqnarray}
Then, it can be shown that  with $f_X(x), \, f_P(p)$, and $\varepsilon^* $ given and with $X_b$ and $P_b$ having the properties described in Assumptions, one can choose $n$ large enough that the complex-valued function $\psi$ given by the construction can be used to uniformly approximate the cumulative distributions $F_X(x)$ and $F_P(p)$ for $x \in [-X_{b}, X_{b}]$, and for $p \in [-P_{b}-\triangle p/2, P_{b}-\triangle p/2]$:

\begin{eqnarray}
\label{claim11}
\left | F_X(x) - \widetilde{F}_X(x) \right | \leq \varepsilon^*
\end{eqnarray}
and
\begin{eqnarray}
\label{claim12}
\left | F_P(p) - \widetilde{F}_P(p) \right | \leq \varepsilon^*
\end{eqnarray}

For derivation of (\ref{claim1}) and (\ref{claim11}) see Appendix. Approximations (\ref{claim2}) and (\ref{claim12}) are derived in a similar fashion.
\section{Examples}
To illustrate the approximation,  we compare the integral of the probability density function over a series of intervals with the correpsonding approximation generated by $\psi$ from (\ref{claim1}), (\ref{claim2}) and (\ref{phidef}). Specifically, in $P$ space we plot
 $\int^{p+\triangle p/2}_{p-\triangle p/2}\, f_{P}(q)dq$
and the corresponding approximation
        $ \int^{p+\triangle p/2}_{p-\triangle p/2}t \cdot \widehat{\psi}(qt) \widehat{\psi}^*(qt)dq$. Similarly, in $X$ space we plot 
 $\int^{x+\triangle x/2}_{x-\triangle x/2}\, f_{X}(q)dq$
and the corresponding approximation
        $ \int^{x+\triangle x/2}_{x-\triangle x/2}
     \psi(q)   \cdot \psi^*(q) dq$; we approximate the last integral by $\psi(x)   \psi^*(x) \triangle x$, as $\triangle x$ is small. 

We applied the method to four pairs of distributions; for each pair, we use a single $\psi$ (and its Fourier transform) to approximate them  with $n$=7, $s$=64 and $\triangle x =\triangle p = 5/2^n \approx 0.039$:
\begin{enumerate}
\item Figure 1 illustrates the approximation using Equation (\ref{phidef}), for the distributions
\begin{align*}
&f_X(x)={4 \over {\sqrt{2\pi}}}\biggl[0.75\,e^{-8\,(x+{1 \over{2}})^2} + 0.25\,e^{-8\,(x-{1 \over{2}})^2}\biggr]\\
&\mathrm{and}\;\;f_P(p)= {1 \over {\sqrt{2\pi}}}e^{-p^2/2}
\end{align*}
\item Figure 2 illustrates the approximation using Equation (\ref{phidef}), for the distributions
\[f_X(x)= e^{-x},\;x\geq 0\;\;\mathrm{and}\;\;f_P(p)= {1 \over {\sqrt{2\pi}}}e^{-p^2/2}\]
\item Figure 3 illustrates the approximation using Equation (\ref{phidef}), for the distributions
\[f_X(x)= {1 \over {\sqrt{2\pi}}}e^{-x^2/2}\;\;\mathrm{and}\;\;f_P(p)= e^{-p},\;p\geq 0\]
\item Figure 4 illustrates the approximation using Equation (\ref{phidef}), for the distributions
\[f_X(x)= {1 \over {\sqrt{2\pi}}}e^{-x^2/2}\;\;\mathrm{and}\;\;f_P(p)= {1 \over {\sqrt{2\pi}}}e^{-p^2/2}\]
\end{enumerate}
\section{Discussion}
Based on the algorithm described in section 3 we have shown how to approximate two given distributions from a single complex-valued function with a specified accuracy. The main result of our paper is given in expressions (\ref{claim1}-\ref{claim2}), (\ref{claim11}-\ref{claim12}) and (\ref{phidef}). Expression (\ref{phidef}) can be viewed as an extension of well-known  numerical methods based on the sinc function  \cite{sinc}, providing a method for approximating two distributions simultaneously.

The construction and arguments given here show that there always exists a function of the form given by (\ref{phidef}) which approximates two given distributions with a specified degree of accuracy provided our assumptions (1-3) are satisfied. For successively higher degrees of accuracy, the method allows construction of a sequence of approximating functions but this sequence may not necessarily converge, say in $L^2$, to a limiting function.

A modification of our approach or a similar method based not on the sinc function but on some other basis or wavelet series \cite{wave} might be used to approximate the two distributions and  converge to a limiting function. For example, one might attempt to use the Hermite polynomials multiplied by the appropriate Gaussian function as a basis.

It remains speculative, however, whether our approach can be modified so as to approximate arbitrary pairs of distributions with a basis other than the sinc functions.
Further speculation about this potential generalization is beyond the scope of this paper. We would like to stress that the main goal of the present work is not to find the best numeric approximation (though we can always meet an increasing demand in accuracy) but to establish the existence of an algorithm allowing uniform approximation of the cumulative distributions.

\renewcommand{\theequation}{A-\arabic{equation}}
  \setcounter{equation}{0}
  \section*{APPENDIX}
  We sketch the derivation of (\ref{claim1}) and (\ref{claim11}). 
First we write down the Fourier transform of $\psi(x)$:
\begin{eqnarray}
\label{Ftrans}
  \widehat{\psi}(q)&=&\int \psi (x) e^{-i\cdot 2\pi \cdot x\cdot q} dx \cr\cr
   & =& \sqrt {\frac{s}{\triangle x}}\int \sum ^{sJ-1}_{j=-sJ} a_j e^{-i\cdot 2\pi \cdot M_j\cdot x}\cr\cr
   &&\textrm{sinc}(  \frac{x-j \triangle x /s}{\triangle x /s}  ) e^{-i\cdot 2\pi \cdot x\cdot q} dx \cr\cr
   & =& \sqrt {\frac{\triangle x}{s}} \sum ^{sJ-1}_{j=-sJ} a_j R\biggl(\frac{\triangle x}{s}(q-M_j) \biggr)\cr\cr 
   &&e^{-i\cdot
   2\pi \cdot j\frac{\triangle x}{s}(q-M_j)}
\end{eqnarray}
In (\ref{Ftrans}) $R$ is a rectangular function:
\begin{equation}
\label{rr}
R(q)=\begin{cases} 1& \text{when $-{1\over 2}<q<{1\over 2}$},\\ {1\over 2}& \text{when $q=\pm{1\over 2}$}, \\ 0& \text{otherwise}
\end{cases}
\end{equation}
We consider
\begin{equation}
\label{II}
I(l)\,\equiv\,\int^{(l+{1\over{2}})\triangle p}_{(l-{1\over{2}})\triangle p}t \cdot \widehat{\psi}(qt)   \cdot
  \widehat{\psi}^*(qt)dq,
\end{equation}
the integral of $t \cdot \widehat{\psi}(qt)   \cdot \widehat{\psi}^*(qt)$ over the interval $[(l-{1\over{2}})\triangle p,\; (l+{1\over{2}})\triangle p]$, for each $l \in {-J, -J+1,...,
J- 1}$. Using  expression (\ref{Ftrans}), rescaling the integration variable, and using definition of integer $m_{j}:\;m_{j}=\triangle xM_{j}/s$, for this integral we obtain:
\begin{eqnarray}
\nonumber
I(l)&=&\int^{(l+{1\over{2}})}_{(l-{1\over{2}})}dq^{\prime}
  \sum ^{sJ-1}_{j,k=-sJ}a_ja_kR(q^{\prime}-m_j)R(q^{\prime}-m_k)\cr\cr
  &&e^{- 2i\pi
   ((j-k)q^{\prime}+(j\cdot m_j-k\cdot m_k))}
\end{eqnarray}
where $a_j = [\triangle x f_X(j\triangle x/s)/s]^{1/2}$.

Property (\ref{rr}) of the rectangular function $R$ implies that each $j,k$ term in the integrand is either 0 for the full range
 of integration or vanishes for all but a subinterval of length 1. The latter holds only if $m_j = m_k$ and we
 have: $l \leq  m_j \pm \frac{1}{2} \leq (l+1)$ if and only if  $l = m_j$, for
 some integer $m_j$.  After interchanging the order of integration and summation, each integral in the
 sum is 0 due to the factor $e^{i\cdot 2\pi \cdot(j-k)\cdot q^{\prime}}$, unless $j = k$. On the other hand, if
 $j = k$ the integral is just $a_j ^2$.  Thus, we can write ($l=-J,-J+1,...,J-1$):
\begin{eqnarray}
\label{F trans1}
I(l)=\int^{(l+{1\over{2}})\triangle p}_{(l-{1\over{2}})\triangle p} t \cdot \widehat{\psi}(qt)   \cdot
\widehat{\psi}^*(qt)dq = \sum _{j \in S_l} a_j^2,
\end{eqnarray}
where we recall the definition of $S_l$: $S_l$ is the set of all $j \in {-sJ, -sJ+1,...,sJ}$ for which
$M_j\cdot (\triangle x /s)= l$.

There are $2J$ terms $I(l)$ of the form given in (\ref{F trans1}); the approximation given by inequality (\ref{claim1}) will be proven if we can show that for $j=-J, -J+1,...,J-1$ the following relation holds:
\begin{eqnarray}
\label{F trans2}
\left|{\sum_{j\in S_l} a_j^2-pr_{p;l}}\right|&=&\sum_{j\in S_l} \left|{ a_j^2- \int^{(l+{1\over{2}})\triangle
 p}_{(l-{1\over{2}})\triangle p} f_P(z)dz}\right|\cr\cr
 &&\leq\,{\varepsilon\over 2J}
\end{eqnarray}
Either inequality (\ref{F trans2}) holds for $l =-J, -J+1,...,J-1$, and we are done, or there is a
smallest $l$ violating (\ref{F trans2}), say $l_1$, such that $\sum_{j\in S_{l_1}}| a_j^2- \int^{(l+{1\over{2}})\triangle p}_{(l-{1\over{2}})\triangle p} f_p(z)dx| \\> \varepsilon/(2J) $. This implies that $j_1 \geq sJ-1$,
where $j_1 = max \{j: j \in S_l\}$.  Indeed, if this were not the case (i.e. $j_1 < sJ-1$), than $M_{j_1+1}$ would have been defined by the algorithm described in  section 3 as
$l_1$ since $a_{j_{1}+1} \leq \varepsilon/(2J+1)$ and $j_1+1$ would also have been in $S_{l_1}$. This would
be a contradiction since $j_1$ was supposed to be the largest element of $S_{l_1}$. Therefore, $j_1 \geq sJ-1$.

By choice of $s$ in section 3, requirement (\ref{c4}): $ |\sum ^{J-1}_{l=-J} \sum_{j\in S_l} a_j^2-
\int^{(j+1)\triangle x/s}_{j\triangle x/s} f_X(x)dx |$ is less or equal  $\varepsilon/8,$  and
because $f_X(x)$ integrates over the interval $[-X_b, X_b]$ to at least $1-\varepsilon/16$ (Assumption 3),
we have:
\begin{eqnarray}
\label{F trans3}
\sum ^{sJ-1}_{j=-sJ} a_j^2 \geq 1- \varepsilon/16 -\varepsilon/8 > 1-\varepsilon/4
\end{eqnarray}
Since $f_P(p)$ is a probability density function and by Assumption 3 and the algorithm described in section 3:
\begin{eqnarray}
\label{F trans4}
 1 &\geq& \int ^{(l_1+{1\over{2}})\triangle p}_{-P_b-{1\over{2}}\triangle p}f_P(z)dz= \sum^{l_1}_{l=-J}  \int^{(l+{1\over{2}})\triangle p}_{(l+{1\over{2}})\triangle p} f_P(z)dz \cr
  &\geq& \sum^{l_1}_{l=-J} \sum_{j \in S_l}a_j^2
  = \sum^{sJ-1}_{j=-sJ} a_j^2\geq 1- \varepsilon/4.
\end{eqnarray}
The last inequality above is simply inequality (\ref{F trans3}). By the Construction, $\int^{(l+{1\over{2}})\triangle
p}_{(l-{1\over{2}})\triangle p} f_P(z)dz \geq \sum_{j \in S_l}a_j^2$   for every $l$.  This fact and (\ref{F trans4}) imply:
\begin{eqnarray}
\label{F trans5}
\int ^{(l+{1\over{2}})\triangle p}_{-P_b-{1\over{2}}\triangle p}f_P(z)dz -\sum^{l_1}_{l=-J} \sum_{j \in
S_l}a_j^2 = \cr\cr\cr \sum^{l_1}_{l=-J}\left |{\int ^{(l+{1\over{2}})\triangle p}_{(l-{1\over{2}})\triangle p} f_P(z)dz
-\sum_{j \in S_l}a_j^2}\right | \leq \frac{1}{4}\varepsilon
\end{eqnarray}
Further, the upper bound of 1 in expression (\ref{F trans3}) implies that:
\begin{eqnarray}
\label{F trans6}
  \int ^{P_b-{1\over{2}}\triangle p}_{(l_1+{1\over{2}})\triangle p}f_P(z)dz& =& \sum_{l=l_{1}+1}^{J-1}\int
  ^{(l+{1\over{2}})\triangle p}_{(l-{1\over{2}})\triangle p} f_P(z)dz \cr\cr
  &\leq& \frac{3}{4}\varepsilon
\end{eqnarray}

Now $S_l$ is empty for $l > l_1$ since $M_{sJ-1} = l_1$ and $M_j$ is non-decreasing. In other words $\sum_{j \in S_l}a_j^2 =0$ for $l > l_1$. This fact and the
inequality in Expression (\ref{F trans6}) yield:
\begin{eqnarray}
\label{F trans7}
\left|{\sum_{l=l_1+1}^{J-1}\int ^{(l+1)\triangle p}_{l\triangle p} f_P(z)dz-\sum_{j \in
S_l}a_j^2  }\right|=\cr\cr\cr \left|{\sum_{l=l_1+1}^{J-1}\int ^{(l+{1\over{2}})\triangle p}_{(l-{1\over{2}})\triangle p}
f_P(z)dz-0 }\right|\leq \frac{3}{4}\varepsilon
\end{eqnarray}
Finally, inequalities (\ref{F trans5}) and (\ref{F trans7}) lead to the desired result, proving
inequality (\ref{claim1}).

Proof of inequality (\ref{claim2}) requires no new qualitative features but straightforward tedious calculations; we do not cite it here.

To prove the uniform approximation for the cumulative distributions, inequality (\ref{claim11}), choose $n$ large enough that $Q_x\cdot \triangle x = Q_x\cdot X_b/2^n  < \varepsilon^*/2$, construct $\psi(x)$ so that inequalities (\ref{claim1}) and (\ref{claim2})  hold with $\varepsilon$=$\varepsilon^*/2$ and let $\delta(x)=\left | F_X(x) - \widetilde{F}_X(x) \right |$. We now  show that $\delta(x)  \leq \varepsilon^* $ for all $x \in [j\triangle x, \, (j+1)\triangle x]$ and for $j = -J, -J+1,\cdots, J-1$. Suppose the maximum of $\left | F_X(x) - \widetilde{F}_X(x) \right |$ occurs at $X = x_0$, for $x_0 \in [j\triangle x, \, (j+1)\triangle x]$. Let $M_{F_j}$ denote $Max\{F_X(x), x \in [j\triangle x, \, (j+1)\triangle x]\}$. If  $F_X(x_0) > \widetilde{F}_X(x_0)$, then
\begin{eqnarray}
\label{eq1}
 \delta(x) \leq M_{F_j} - \widetilde{F}_X(j\triangle x)
\end{eqnarray}
since $\widetilde{F}_X(x)$ is non-decreasing. Further, the mean value theorem implies
$M_{F_j} \leq F_X(j\triangle x) + Q_x\cdot \triangle x $ since $Q_x$ is the maximum of $f_X(x)$. Using this result in (\ref{eq1}) gives
 \begin{eqnarray}
\delta(x) &\leq& F_X(j\triangle x) + Q_x\cdot \triangle x - \widetilde{F}_X(j\triangle x) \cr
&\leq& \varepsilon^*/2 + \varepsilon^*/2 = \varepsilon^{*}
\end{eqnarray}
since $\left | F_X(j\triangle x) - \widetilde{F}_X(j\triangle x) \right | \leq \varepsilon^*/2$ follows from (\ref{claim1}) and $| Q_x\cdot \triangle x | \leq \varepsilon^*/2$ by choice of $n$.

On the other hand, if $F_X(x_0) < \widetilde{F}_X(x_0)$, then
\begin{eqnarray}
\delta (x) &\leq& \widetilde{F}_X((j+1)\triangle x)-  F_X(j\triangle x) \cr
&\leq& F_X((j+1)\triangle x)+ \varepsilon^*/2 - F_X(j\triangle x) \cr
&\leq& F_X(j\triangle x) + Q_x\cdot \triangle x + \varepsilon - F_X(j\triangle x) \cr
&\leq& \varepsilon^*/2 + \varepsilon^*/2 = \varepsilon^{*}
\end{eqnarray}
The chain of estimates above follows from the fact that $\widetilde{F}_X(x)$ is non-decreasing, $Q_x$ is the maximum of $f_X(x)$, and by choice of $n$.

The proof for $\left | F_P(p) - \widetilde{F}_P(p) \right |$ is similar.

\newpage
\begin{figure}
\centering
\includegraphics[width=2.55in]{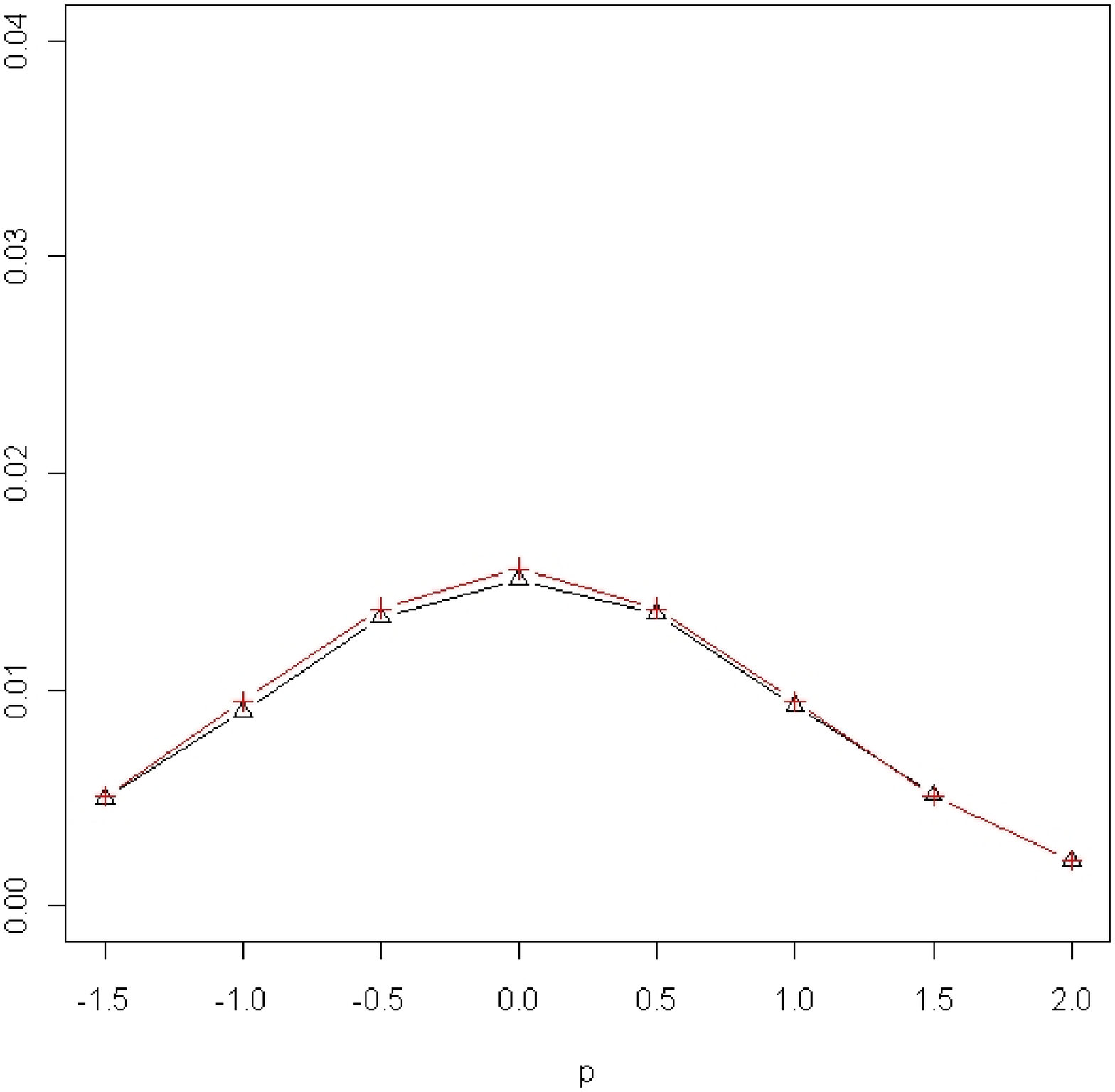}
 \includegraphics[width=2.55in]{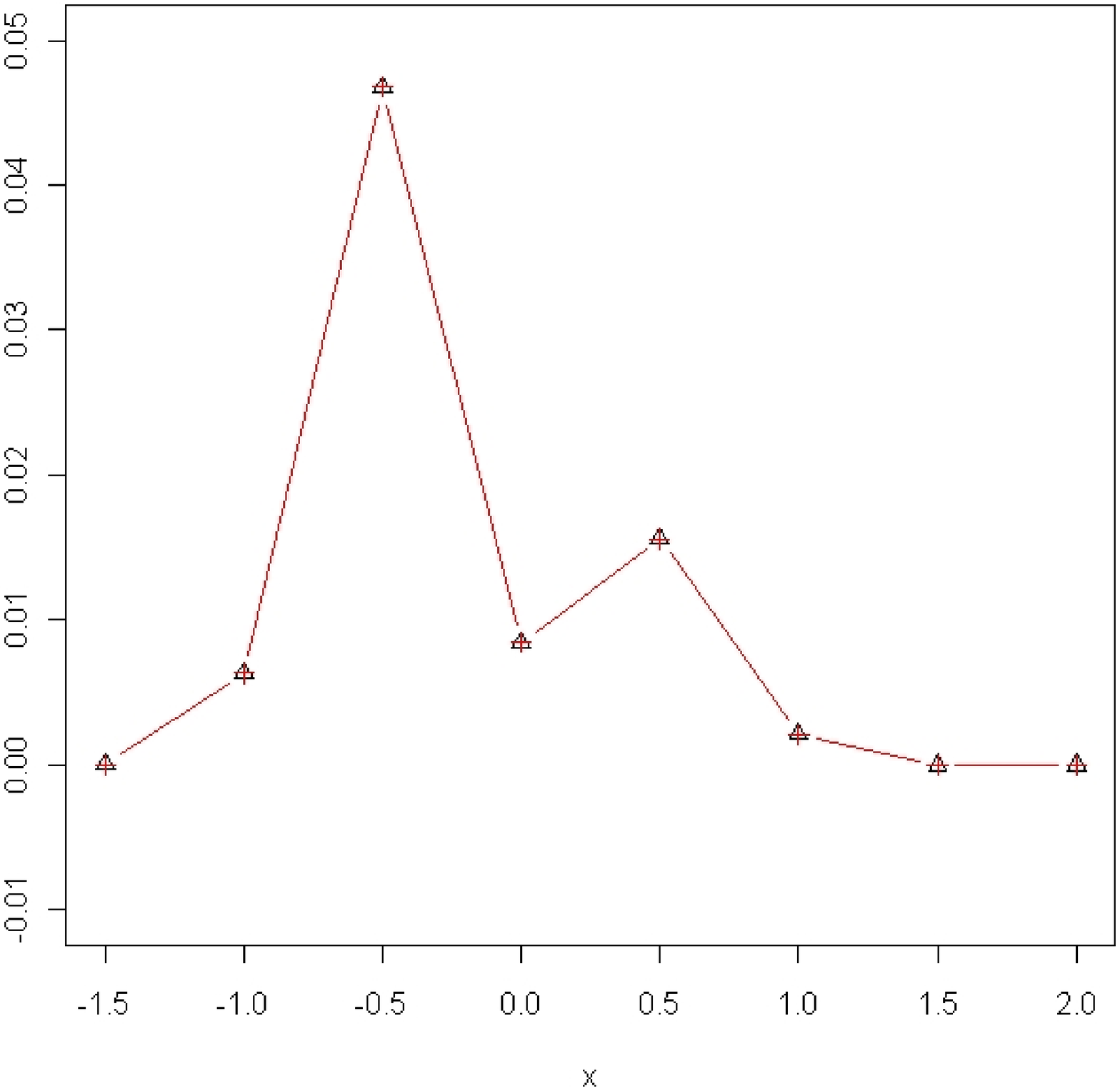}
\caption{Approximation to the integrals $\int^{p+\triangle p/2}_{p-\triangle p/2}\, f_{P}(q)dq$ (top) and 
 $\int^{x+\triangle x/2}_{x-\triangle x/2}\, f_{X}(q)dq$ (bottom) based on Equation (\ref{phidef}) for the pair of distributions $f_{P}(p)\,=\,\exp(-p^{2})/\sqrt{2\pi}$, and $f_X(x)\,=\, {4 \over {\sqrt{2\pi}}}(0.75\,e^{-8\,(x+{1 \over{2}})^2} + 0.25\,e^{-8\,(x-{1 \over{2}})^2} )$.}
\label{1}
\end{figure}
\vspace{400mm}
\begin{figure}
\centering
\includegraphics[width=2.35in]{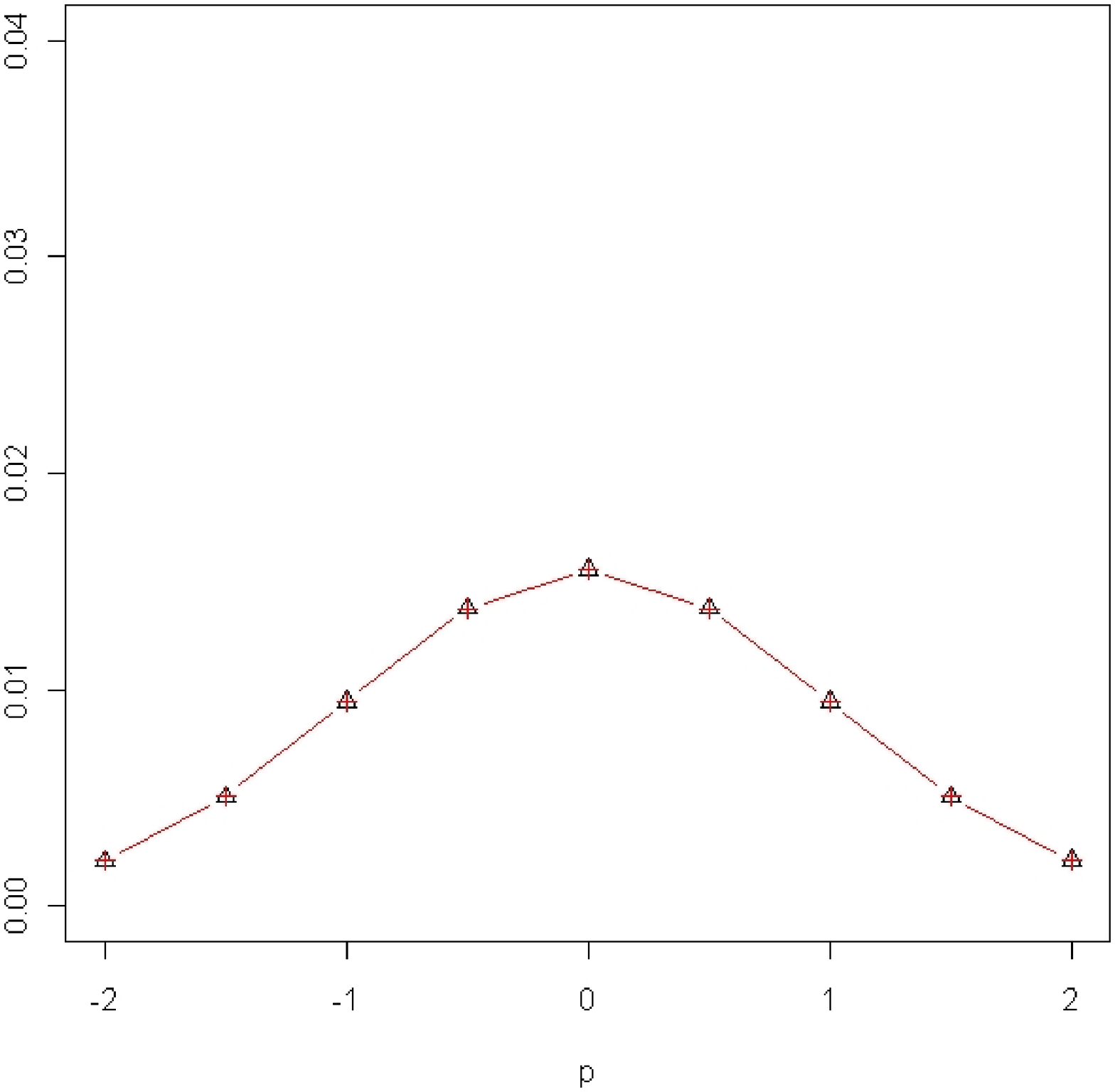}
\includegraphics[width=2.35in]{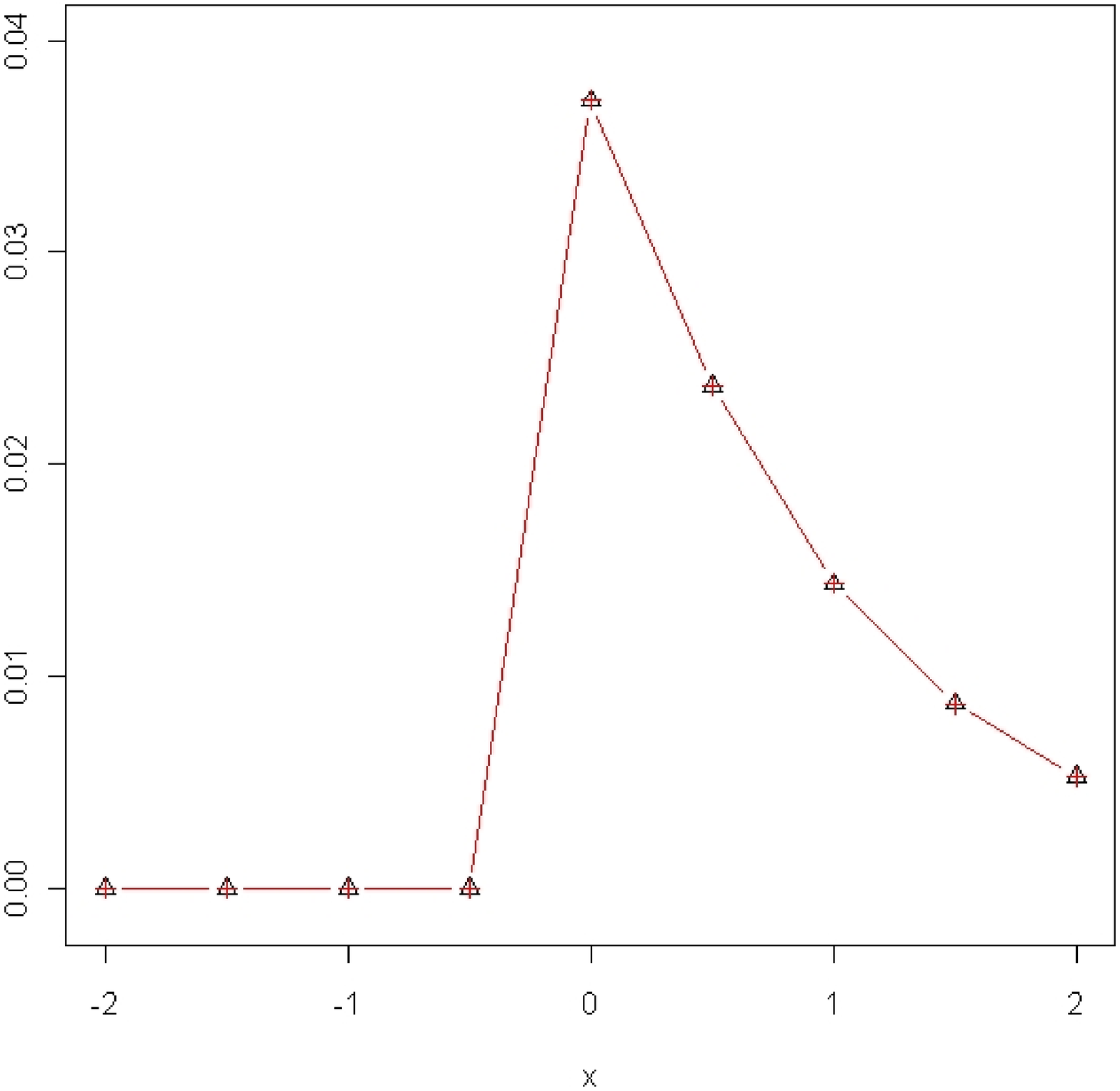}
\caption{Approximation to the integrals $\int^{p+\triangle p/2}_{p-\triangle p/2}\, f_{P}(q)dq$ (top) and 
 $\int^{x+\triangle x/2}_{x-\triangle x/2}\, f_{X}(q)dq$ (bottom) based on Equation (\ref{phidef}) for the pair of distributions $f_P(p)\,=\,{1 \over {\sqrt{2\pi}}}e^{-p^2/2}$, and $f_X(x)\,=\, e^{-x}$ for $x \geq 0$ and $f_X(x)\,=\,0$ for $x<0$.}
\label{1}
\end{figure}
\begin{figure}
\centering
\includegraphics[width=2.35in]{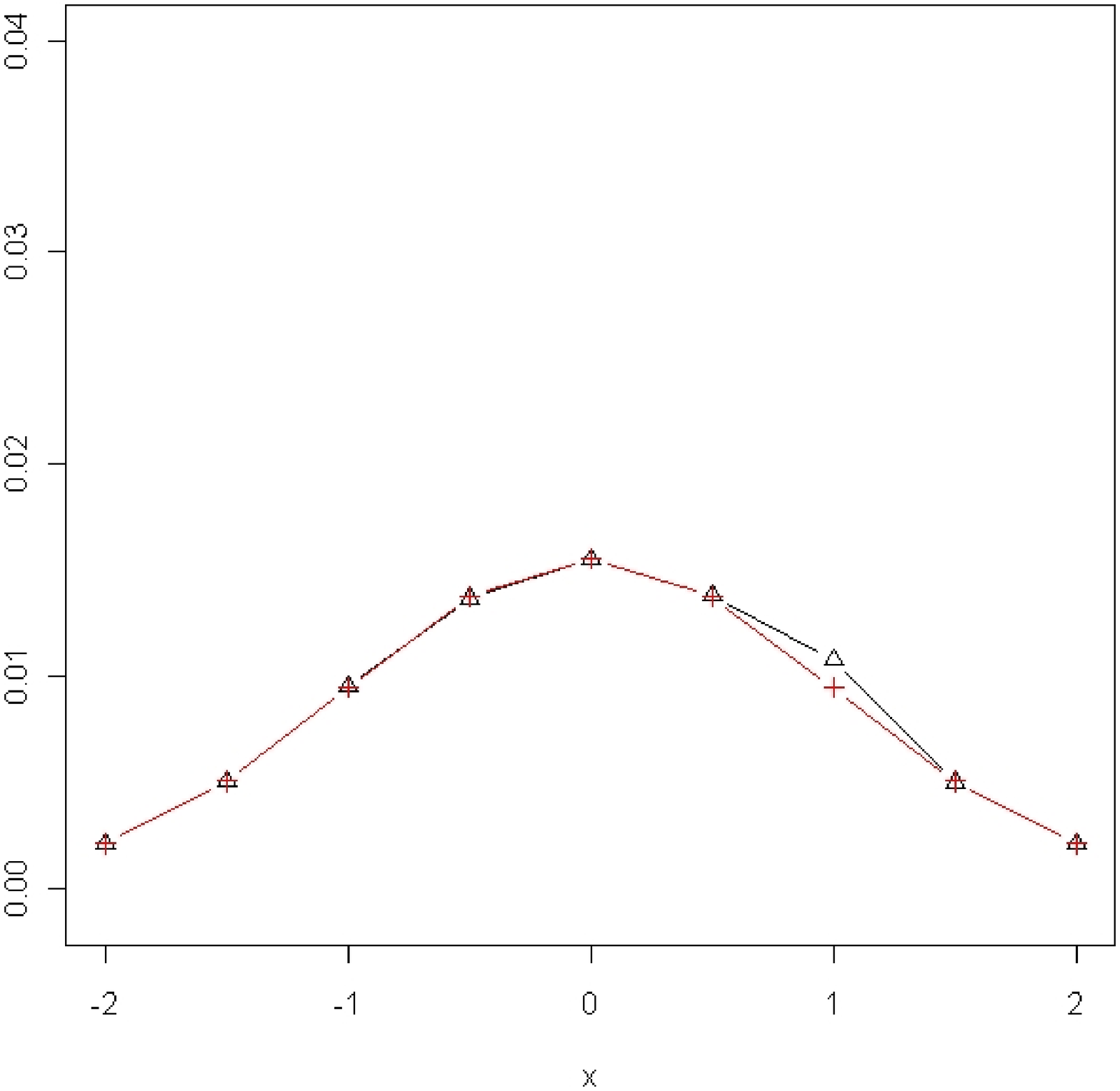}
\includegraphics[width=2.35in]{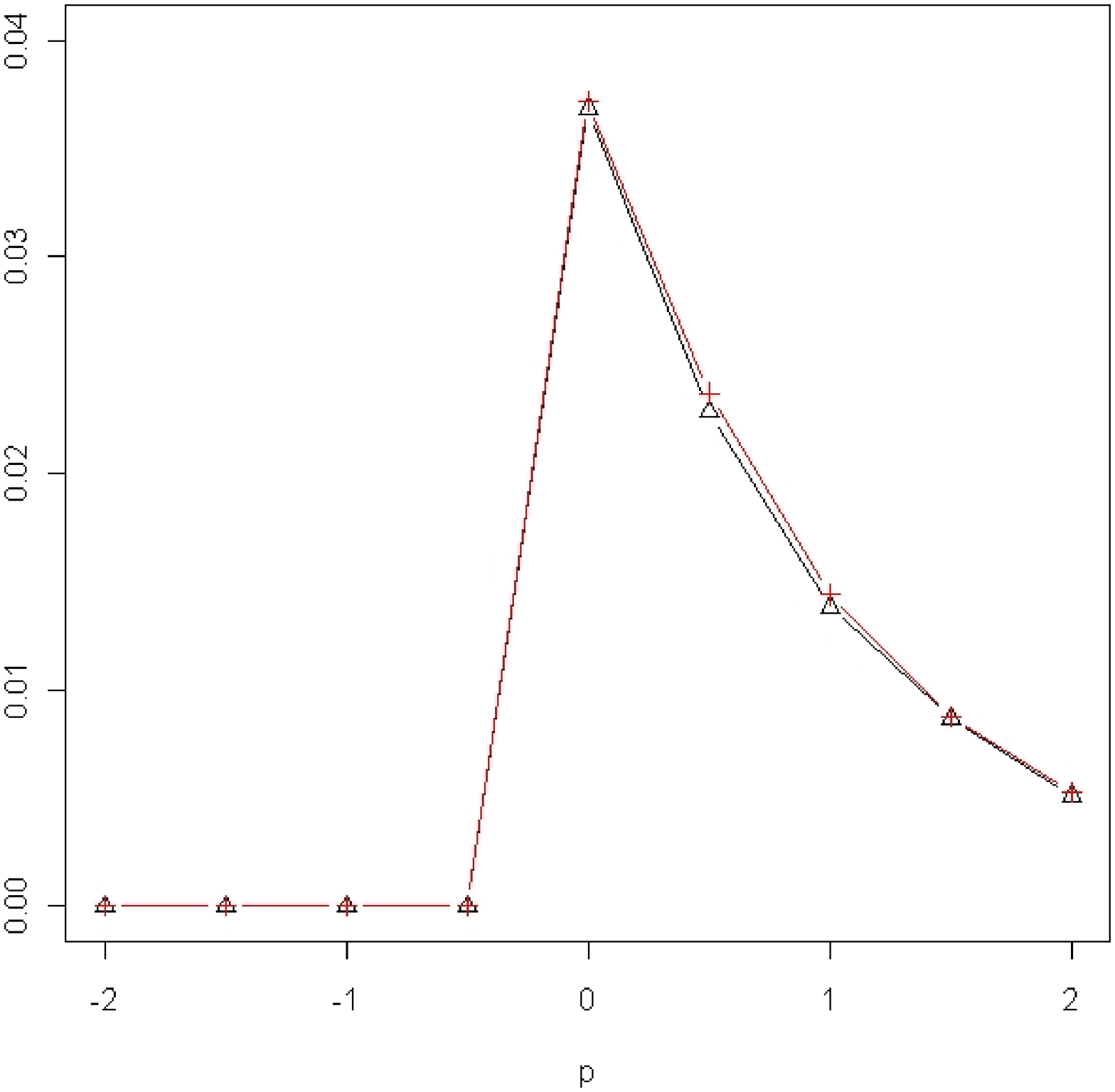}
\caption{Approximation to the integrals $\int^{p+\triangle p/2}_{p-\triangle p/2}\, f_{P}(q)dq$ (bottom) and 
 $\int^{x+\triangle x/2}_{x-\triangle x/2}\, f_{X}(q)dq$ (top) based on Equation (\ref{phidef}) for the pair of distributions $f_X(x)\,=\,{1 \over {\sqrt{2\pi}}}e^{-x^2/2}$, and $f_P(p)\,=\,exp(-p)$ for $p \geq  0$ and $f_P(p)\,=\,0$ for $p < 0$.}
\label{1}
\end{figure}

\begin{figure}
\centering
\includegraphics[width=2.35in]{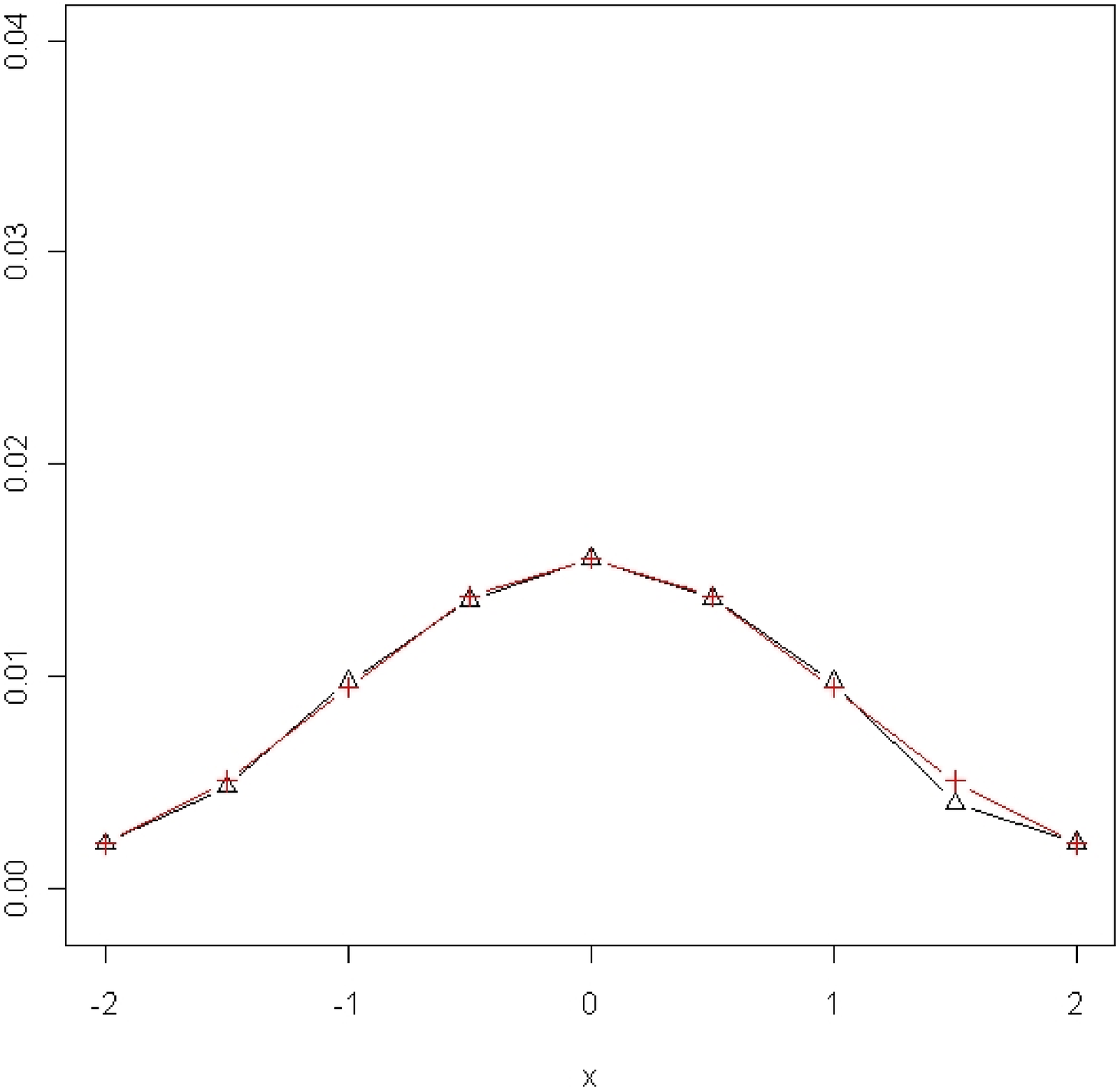}
\includegraphics[width=2.35in]{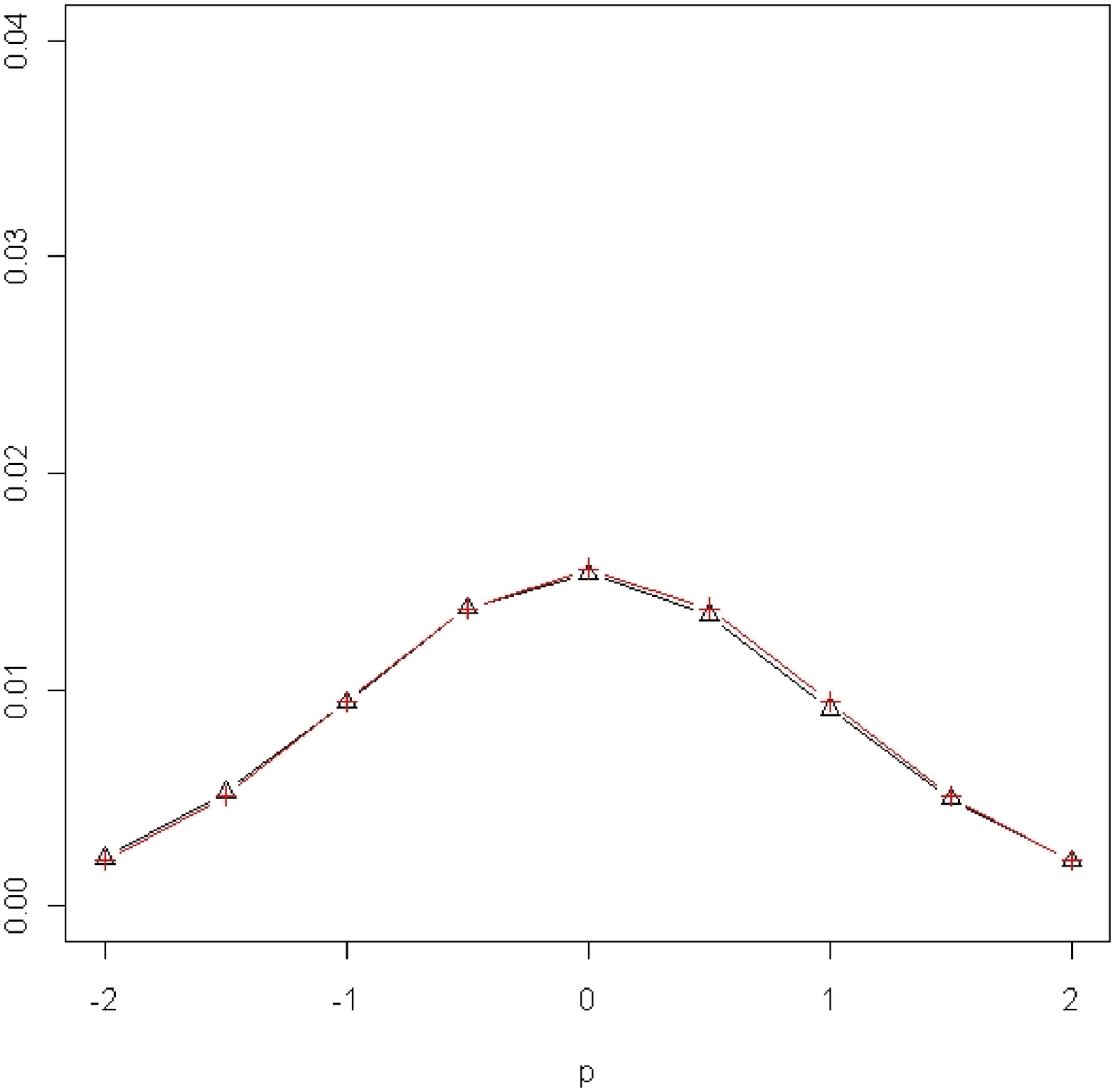}
\caption{Approximation to the integrals $\int^{p+\triangle p/2}_{p-\triangle p/2}\, f_{P}(q)dq$ (bottom) and 
 $\int^{x+\triangle x/2}_{x-\triangle x/2}\, f_{X}(q)dq$ (top) based on Equation (\ref{phidef}) for the pair of distributions $f_P(p)\,=\,{1 \over {\sqrt{2\pi}}}e^{-p^2/2}$, and $f_X(x)\,=\,{1 \over {\sqrt{2\pi}}}e^{-x^2/2}$.}
\label{1}
\end{figure}

\end{document}